%% file: paper.tex
\documentclass[authorversion=true, nonacm=true]{acmart}
\usepackage{listings}
\usepackage{float}
\usepackage{color}
\usepackage{tabularx}
\usepackage{graphicx}

\newif\ifBIBLATEX
\BIBLATEXfalse

\newif\ifDEBUG
\DEBUGfalse

\newif\ifANONYMOUS
\ANONYMOUSfalse

\ifBIBLATEX
  \usepackage[style=numeric,
  maxnames=200,
  backend=biber,
  firstinits=true]{biblatex}
\fi

\usepackage{multirow}

\input{typesetting}

\input{data/data}

\acmConference[SCORED '23]{}{November 2023}{Copenhagen, Denmark}
\acmPrice{15.00}
\acmISBN{XXX}

\acmSubmissionID{scor058}

\makeatother

\begin{document}

\begin{abstract}

As we increasingly depend on software systems, the consequences of breaches in the software supply chain become more severe. High-profile cyber attacks like those on SolarWinds and ShadowHammer have resulted in significant financial and data losses, underlining the need for stronger cybersecurity. One way to prevent future breaches is by studying past failures. However, traditional methods of analyzing these failures require manually reading and summarizing reports about them. Automated support could reduce costs and allow analysis of more failures. 
Natural Language Processing (NLP) techniques such as Large Language Models (LLMs) could be leveraged to assist the analysis of failures.

In this study, we assessed the ability of Large Language Models (LLMs) to analyze historical software supply chain breaches. 
We used LLMs to replicate the manual analysis of 69 software supply chain security failures performed by members of the Cloud Native Computing Foundation (CNCF).
We developed prompts for LLMs to categorize these by four dimensions: type of compromise, intent, nature, and impact. 
GPT 3.5's categorizations had an average accuracy of 68\% and Bard's had an accuracy of 58\% over these dimensions. 
We report that LLMs effectively characterize software supply chain failures when the source articles are detailed enough for consensus among manual analysts, but cannot yet replace human analysts.
Future work can
  improve LLM performance in this context,
  and
  study a broader range of articles and failures.



\end{abstract}




\ifANONYMOUS
\author{Anonymous author(s)}
\else
\author{Tanmay Singla}
\orcid{}
\affiliation{%
  \institution{Purdue University}
  \country{West Lafayette, IN, USA}}
\email{singlat@purdue.edu}

\author{Dharun Anandayuvaraj }
\orcid{}
\affiliation{%
  \institution{Purdue University}
  \country{West Lafayette, IN, USA}}
\email{pdananday@purdue.edu}

\author{Kelechi G. Kalu}
\orcid{}
\affiliation{%
  \institution{Purdue University}
  \country{West Lafayette, IN, USA}}
\email{kalu@purdue.edu}

\author{Taylor R. Schorlemmer}
\orcid{}
\affiliation{%
  \institution{Purdue University}
  \country{West Lafayette, IN, USA}}
\email{tschorle@purdue.edu}

\author{James C. Davis}
\orcid{0000-0003-2495-686X}
\affiliation{%
  \institution{Purdue University}
  \country{West Lafayette, IN, USA}}
\email{davisjam@purdue.edu}

\renewcommand{\shortauthors}{Singla \etal}
\fi


\newcommand{\MyTitle}{}
\renewcommand{\MyTitle}

\renewcommand{\MyTitle}{Large Language Models to Analyze Software Supply Chain Security Incidents}
\renewcommand{\MyTitle}{Evaluating the Efficacy of Large Language Models in the Analysis of Software Supply Chain Security Incidents}
\renewcommand{\MyTitle}{An Empirical Study on Using Large Language Models to Analyze Software Supply Chain Security Failures}

\title{\MyTitle}

\maketitle

\section{Introduction}

\TODO{Need to use consistent terms: incidents/failures/breaches, analyze/study/labeling...}
\JD{Recommend we use ``failures'' throughout, but in \$2 (maybe \$2.3?) we should put in definitions of defect, failure, incident, and then in 4.1 we need to say these are failures because not all actually led to incidents}
Software mediates almost all aspects of modern life~\cite{jones2011economics}.
To reduce development time, software applications integrate dependencies both directly (\eg importing a library) and indirectly (\eg that library's dependencies).
These dependencies may come to dominate the application's risk profile: it has been estimated that the source code of a typical web application is comprised of 80\% dependencies and only 20\% custom business logic~\cite{pashchenko_vulnerable_2018, noauthor_2023_nodate}.
The owners of these dependencies may be external to the organization developing the application, and thus the reduction of development time comes with an increase in risks associated with this \emph{software supply chain}~\cite{ellison_evaluating_2010,gokkaya_software_2023}.
One potential risk is a \emph{software supply chain attack} --- actors insert or exploit vulnerable logic in dependencies, these dependencies are integrated into applications, and the vulnerability becomes exploitable in application deployments~\cite{sok2022}.

\begin{figure}[t]
    \centering
\includegraphics[bb=0 0 200 109,width=0.8\textwidth]{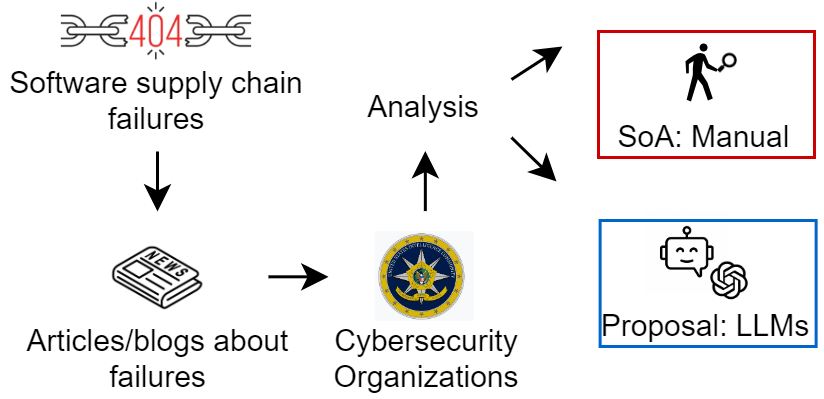}
\caption{
    Proposed use of Large Language Models (LLMs) to analyze software supply chain failures.
    Failures are often reported in articles and blogs.
    Organizations concerned with cybersecurity (\emph{e.g.} governments, corporations) manually analyze failure reports.
    We evaluate LLMs as an aid.
    }
    \label{fig:MethodOverview}
\end{figure}
\TS{Fig 1 “Proposal on LLMs” is to close to left blue line
Fig 2 weird warping?
Fig 3 I think the arrows are a bit small? Could look more pleasing
Fig 1-3 could be fixed by adjusting white spaces, and also in general I think rounded boxes and curves are just more aesthetically pleasing (but again just aesthetics that are easy to change)}

In a failure-aware engineering process, engineers study past failures to prevent future ones~\cite{petroski1994design,anandayuvaraj2023incorporating}.
Although organizations may be unwilling to publicly disclose their own failures, news articles and other kinds of grey literature could provide sufficient information on failures~\cite{anandayuvaraj_reflecting_2023}.
Such data comprises ``Open-Source Intelligence''~\cite{recordedfuture2022}, and are used by governmental bodies, military institutions, and law enforcement agencies~\cite{gill2023} to design security offenses and defenses. 


Current approaches to garnering open-source intelligence, \eg studying news articles of failures, require costly expert manual analysis.
For example, the Cloud Native Computing Foundation (CNCF) maintains a database of software supply chain security failures through manual analysis~\cite{CNCF2023}.
This database has been further analyzed manually~\cite{geer2020good}. 
With the goal of reducing the costs of manual analysis, we assess the effectiveness of Large Language Models (LLMs) in gathering open-source intelligence.
We explored the effectiveness of LLMs at replicating the classifications of the CNCF database~\cite{CNCF2023} made by Geer \etal~\cite{geer2020good} and the CNCF database maintainers.
We conducted prompt engineering to iteratively develop prompts that performed well on a sample of 20\% of the articles and then evaluated performance on the remaining 80\%.
In addition, we introduced a new category of analysis, ``Lessons learned'', to assess the usefulness of an LLM's recommendations.

\JD{@Tanmay @Dharun Someone check this, I edited it to have a bit more of a message}
We compared the performance of two state-of-the-art LLMs, OpenAI's GPT and Google's Bard, on these prompts.
GPT outperformed Bard in all cases.
GPT's accuracy ranged from 52-88\% on the pre-defined dimensions.
On the open-ended ``Lessons learned'', our research team rated GPT's performance as reasonable but not excellent, with an average helpfulness score of 3.83/5.
Not surprisingly, the quality of the LLMs' outputs depends on the level of detail provided in the source articles --- more comprehensive articles lead to higher-quality responses, as well as less disagreement among the manual raters.
Lastly, we note that sometimes we preferred GPT's rating over that provided by the CNCF, suggesting that ground truth may be difficult to establish in this context.

Our contributions are:
\begin{itemize}
  \item An extended analysis of a catalog of software supply chain failures
  \item An evaluation of LLMs at replicating manual characterization of software supply chain failures
  \item An evaluation of LLMs at extracting lessons learned from software supply chain failures
\end{itemize}

\section{Background and Related Work} \label{sec:background}

\subsection{Software Supply Chain} \label{sec:background-supplychain}

Over the years, software production has changed significantly.
Early software engineers wrote most code from scratch, increasing production costs~\cite{vivek_is_2022}.
As reusable libraries and frameworks became more available, software engineers shifted to more software reuse~\cite{sonatype-sotssc}.
Software applications now commonly rely on external code components, often referred to as \emph{dependencies}.
These dependencies, including packages, libraries, frameworks, and other artifacts, serve as building blocks in modern software development~\cite{sonatype-sotssc}.

\begin{figure}[h!]
    \centering
    \includegraphics[bb=0 0 704 404,width=0.8\textwidth]{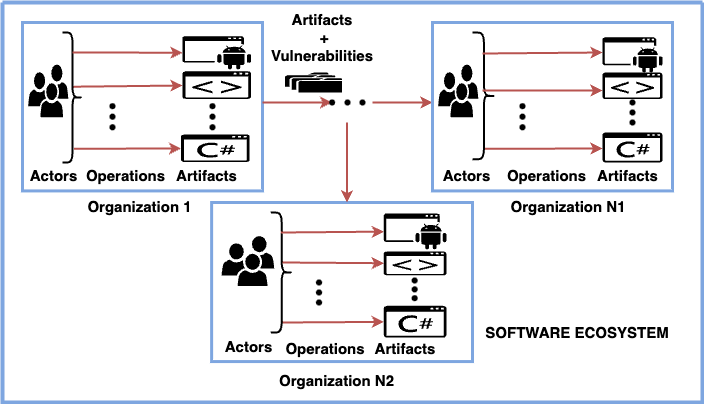}
    \caption{
    A Software Ecosystem's Supply Chain Component and Dependency Vulnerability Flow.
    }
    \label{fig:SoftwareSupplyChain}
\end{figure}

This paradigm shift leads to \emph{software supply chain}: the collection of systems, devices, and people which result in a final software product~\cite{cybersecurity_enisa_2021}.
\cref{fig:SoftwareSupplyChain} provides an illustration.
According to Google~\cite{GoogleCloud2023}, the constituents of a software supply chain include:
 (1) The code developed by teams, its dependencies, and the various internal and external software applications utilized in the development, compilation, packaging, and installation of the software;
 (2) The rules and procedures used in all stages of the process;
 and
 (3) The systems used for the development of the software and its dependencies.
A software supply chain can also be viewed as a network linking \emph{actors} who perform \emph{operations} on \emph{artifacts} ~\cite{ellison_evaluating_2010, nissen2018deliver, sok2022}.

The popularity and reliance on third-party dependencies have been reported in various studies. For example, a 2012 study by Nikiforakis \etal~\cite{nikiforakis_you_2012} showed that 88\% of the Alexa top 10,000 websites included at least one remote JavaScript library. 
Also, according to a 2019 Synopsys Black Duck report, over 96\% of the applications they analyzed include some OSS libraries. These libraries often make up more than 50\% of the average code-base  ~\cite{, ponta_detection_2020}. In the 2023 version of this report, the percentage of code in codebases that was open source had risen to about 80\% ~\cite{pashchenko_vulnerable_2018, noauthor_2023_nodate}.

Software supply chains come with a tradeoff.
Costs are reduced during product development and maintenance, but harm may result due to a mismatch between the desired integrity level of a product and the integrity level achieved by one's dependencies.
Defects in dependencies may cause an application to fail, as we discuss next.

\subsection{Software Supply Chain Attacks} \label{sec:SE-SC-Failures}


Faults in software supply chains leave applications vulnerable to attack~\cite{vasilakis_supply-chain_2021}.
Attacks on software supply chains (or records about them) are a recent trend, following the industry shift to relying on third-party components (\cref{sec:background-supplychain}).
According to a 2021 Sonatype report~\cite{sonatype-sotssc}, from February 2015--June 2019 only 216 software supply chain attacks were recorded, then from July 2019 to May 2020 there were 929 attacks recorded, and from 2020-2021 there were over 12,000 attacks recorded.
In their 2022 report, this number skyrocketed to 88,000~\cite{sonatype_state_2022}.
Some high-profile attacks, such as SolarWinds~\cite{huddleston2021vmware} and ShadowHammer~\cite{kaspersky2019shadowhammer}, threatened US national security.

These and similar attacks have inspired comments from many organizations.
Governmental organizations such as the Cybersecurity and Infrastructure Security Agency (CISA), the National Security Agency (NSA), and the European Union Agency for Cybersecurity (ENISA) have published threat reports and guidance for securing software supply chains~\cite{threat_landscape_enisa_2021, threat_landscape_enisa_2022, threat_landscape_supply_chain_enisa_2021, good_practices_enisa_2023, supply_chain_recommended_practices}.
Industry organizations such as the Cloud Native Computing Foundation (CNCF) have also published their own findings and suggestions~\cite{cloud_native_computing_foundation_software_2021}.
These findings have led to the development of security frameworks such as the widely-recognized Supply-chain Levels for Software Artifacts (SLSA)~\cite{slsa}.

Academics have also begun to focus on software supply chain attacks.
Ohm \etal~\cite{ohm_backstabbers_2020}, Ladisa \etal~\cite{ladisa_taxonomy_2022}, Zimmerman \etal~\cite{Zimmermann2019SecurityThreatsinNPMEcosystem}, and Zahan \etal~\cite{Zahan2022WeakLinksinNPMSupplyChain} studied and characterized attacks on the software supply chain.
Okafor \etal~\cite{sok2022} condensed existing knowledge about software supply chain attacks into a four-stage attack pattern consisting of initial compromise, alteration, propagation, and exploitation.
\cref{table:TypesOfAttacks} summarizes many avenues for these attacks.

{
\begin{table*}[ht]
\small
\centering
\caption{
  Types of software supply chain attacks, according to the Cloud Native Computing Foundation (CNCF)~\cite{CNCF2023}. 
  }
\begin{tabularx}{\textwidth}{ccX}
\toprule
\textbf{ID} & \textbf{Type of compromise} & \textbf{Definition from the catalog} \\
\toprule
1 & Dev Tooling	 & Occurs when the development machine, SDK, tool chains, or build kit have been exploited. These exploits often result in the introduction of a backdoor by an attacker to own the development environment. \\
\midrule
2 &  Negligence & Occurs due to a lack of adherence to best practices. TypoSquatting attacks are a common type of attack associated with negligence, such as when a developer fails to verify the requested dependency name was correct (spelling, name components, glyphs in use, etc). \\
\midrule
3 & Publishing Infrastructure & Occurs when the integrity or availability of shipment, publishing, or distribution mechanisms and infrastructure are affected. This can result from a number of attacks that permit access to the infrastructure.\\
\midrule
4 & Source Code & Occurs when a source code repository (public or private) is manipulated intentionally by the developer or through a developer or repository credential compromise. Source Code compromise can also occur with intentional introduction of security backdoors and bugs in Open Source code contributions by malicious actors.\\
\midrule
5 & Trust and Signing & Occurs when the signing key used is compromised, resulting in a breach of trust of the software from the open source community or software vendor. This kind of compromise results in the legitimate software being replaced with a malicious, modified version.\\
\midrule
6 & Malicious Maintainer & Occurs when a maintainer, or an entity posing as a maintainer, deliberately injects a vulnerability somewhere in the supply chain or in the source code. This kind of compromise could have great consequences because usually the individual executing the attack is considered trustworthy by many. This category includes attacks from experienced maintainers going rogue, account compromise, and new personas performing an attack soon after they have acquired responsibilities. \\
\midrule
7 & Attack Chaining & Sometimes a breach may be attributed to multiple lapses, with several compromises chained together to enable the attack. The attack chain may include types of supply chain attacks as defined here. However, catalogued attack chains often include other types of compromise, such as social engineering or a lack of adherence to best practices for securing publicly accessible infrastructure components.\\
\bottomrule
\end{tabularx}
\label{table:TypesOfAttacks}
\end{table*}
}

\subsection{Failure Studies in Software Engineering} \label{sec:Background-FailureStudies}

Software engineers have finite resources to produce software~\cite{sommerville2015software}.
Engineers accept some defects~\cite{kuutilaTimePressureSoftware2020,costello1984software}, but try to eliminate severe defects that may cause \emph{incidents}: undesired, unplanned, software-induced events that incur substantial loss~\cite{leveson1995safeware}.
Whether severe defects are caught internally or result in incidents, their presence is a \emph{\ul{failure}} indicating a flawed software engineering process.


All engineered systems will fail, regardless of the process (\eg Agile or Plan-based) and methods (\eg test-driven development or formal methods).
For example, Fonseca \etal identified 16 defects across three formally verified systems~\cite{fonsecaEmpiricalStudyCorrectness2017} due to invalid assumptions about the software environment.
Across all schools of software engineering thought, from ISO to Agile, guidelines agree that software engineers should analyze failures to improve for next time~\cite{collier1996defined,basili1993experience,fagan1977inspecting,fagan1999design,gilb1993software,IEEEStandardSoftware2014,ISO9001,ISO90003,kenschwaberScrumGuide2020,beck2000extreme}.
In light of this, techniques to learn from failures~\cite{chenIntelligentIncidentManagement2020} as well as to manage the resulting knowledge~\cite{dingsoyrWhatWeKnow2009} are important software engineering knowledge.

Many researchers have studied software failures in an effort to learn from them~\cite{5718996, geer2020good, noauthor_defending_nodate, anandayuvaraj_reflecting_2023}.
This failure analysis research has advanced the software engineering field~\cite{national2007software, leveson1995safeware, anandayuvaraj_reflecting_2023}. 
However, the high costs associated with failure analysis methods --- which rely on manual analysis --- deter many organizations from undertaking failure analysis~\cite{f8e25d20864a11de931d000ea68e967b}.
In their literature review, Amusuo \etal noted that the typical methodology of academic failure analysis is also manual analysis, and recommended the evaluation of Natural Language Processing (NLP) tools to assist in these tasks~\cite{amusuo_reflections_2022}.
Our study responds by evaluating NLP tools in the context of analyzing cybersecurity failures in the software supply chain.

\subsection{Natural Language Processing in Support of Software Engineering} 

\subsubsection{NLP to Analyze Supply Chain Failures}
In~\cref{sec:SE-SC-Failures} we noted that many governments, companies, and academics are studying software supply chain failures.
To the best of our knowledge, these studies are conducted manually.
This reduces the number of organizations that can gather such intelligence, and we expect that manual efforts will not scale as the number of software supply chain attacks continues to increase.

We believe that recent progress in NLP (Natural Language Processing) could enable large-scale analysis of supply chain failures.
Specifically, recent advancements in \emph{Large Language Models (LLMs)} could aid in studying supply chain failures.
LLMs are neural network-based language models that are capable of ``understanding'' natural language and extracting structured information from unstructured text data~\cite{brantsLargeLanguageModels}.
We therefore hypothesize they could extract relevant failure information from software supply chain failure data sources.
We are not aware of prior work on this topic.

\subsubsection{Other Applications of NLP in SE}
Natural Language Processing (NLP) has been leveraged for various phases of the Software Development Life-Cycle (SDLC).  
NLP tools have been proposed for detecting, extracting, modeling, tracing, classifying, and searching tasks in the specification phase~\cite{zhaoNaturalLanguageProcessing2022}.
NLP tools have been proposed for modeling software systems during the design phase~\cite{bajwaObjectOrientedSoftware}.
NLP tools have been proposed to assist with the development phase by helping detect vulnerabilities and generating code~\cite{ernstNaturalLanguageProgramming2017}.
NLP tools have been proposed to assist during the testing phase~\cite{garousiNLPassistedSoftwareTesting2020}.
NLP tools have been proposed to identify risks during the deployment phase~\cite{vijayakumarAutomatedRiskIdentification2017}.
NLP tools have been proposed to classify user feedback to assist during the maintenance phase~\cite{panichellaHowCanImprove2015}. 
In this paper, we apply NLP tools to learn from failures.

%
%



\section{Research Questions} \label{sec:RQ}

To reduce the costs of analyzing software supply chain failures, we explore the effectiveness of Large Language Models (LLMs) in automating the analysis of these failures.
Towards this goal, we used LLMs to replicate a manual study of software supply chain failures~\cite{CNCF2023}.
Specifically, we investigate:

\begin{itemize}
\item \textbf{RQ1}: How effective are LLMs in replicating manual analysis of software supply chain failures?\\
\item \textbf{RQ2}: Do LLMs suggest viable mitigation strategies for preventing future failures? 
\end{itemize}


\section{Methodology} \label{sec:Methods}
\begin{figure}[h!]

    \centering  
    \includegraphics[bb=0 0 200 150,width=0.7\textwidth]{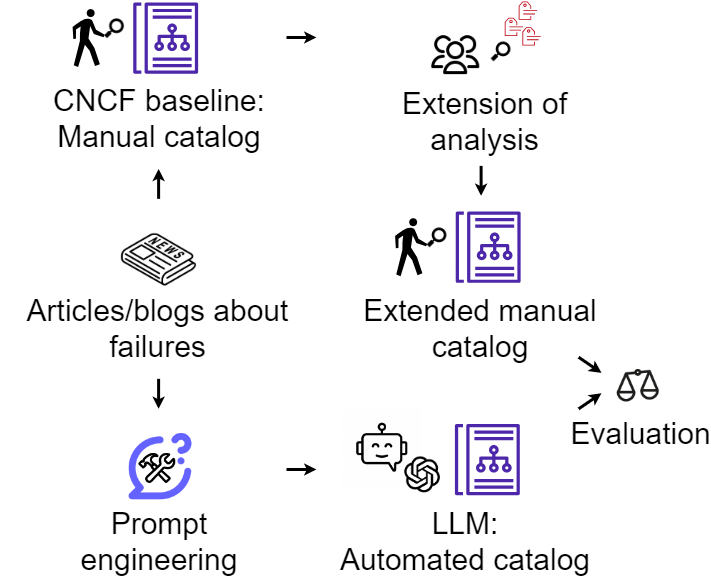}
        \vspace{0.1cm}
    \caption{
    Overview of experiment design. The CNCF catalog manually characterizes software supply chain failures from the news and blogs. We extended this catalog with additional characteristics. We conducted prompt engineering to leverage LLMs to automatically analyze the news and blogs. We compare an LLM's analysis against the manual analysis.
    }
    \vspace{-0.30cm}
    \label{fig:MethodOverview}
\end{figure}

An overview of our methodology is illustrated in~\cref{fig:MethodOverview}.
To assess the effectiveness of LLMs at replicating manual analysis of software supply chain failures, we compare the analysis of a manually generated catalog against the responses generated by two popular LLMs: ChatGPT \cite{chatgptcite} and Bard \cite{bard}.
Specifically to replicate the catalog, we engineered prompts for the LLMs to extract type of compromise, intent, nature, and impact information from the source blogs and news reports. 
Additionally, we constructed a prompt to gather lessons learned, similar to a postmortem~\cite{avizienis2004basic}.
We evaluate the LLM generated catalog for correctness against the CNCF baseline manual catalog. 
We manually extract the intent, nature, and impact information and compare against the LLM's extraction, to evaluate the LLM's effectiveness at conducting an extended failure analysis. 

{
\begin{table*}[ht]
\small
\centering
\caption{
Failure classification examples from CNCF catalog and LLMs. 
}
\begin{tabularx}{\textwidth}{ cp{0.3\textwidth}p{0.2\textwidth}|p{0.2\textwidth}X}
\toprule
\textbf{ID} & \textbf{Name} & \textbf{CNCF's Assessment} &\textbf{GPT 3.5's Assessment} &\textbf{Bard's Assessment} \\
\toprule
1 & RubyGems Package Overwrite Flaw	 & Publishing Infrastructure & Publishing Infrastructure & Publishing Infrastructure\\
2 & Legitimate software update mechanism abused to deliver wiper malware & Publishing Infrastructure & Publishing Infrastructure  & Trust and signing\\
5 & Dropbox GitHub compromise & Attack Chaining & Attack Chaining & Attack Chaining \\
\bottomrule
\end{tabularx}
\label{table:SampleIncidentsAndResults}
\end{table*}
}

\subsection{Articles for analysis} 
The CNCF's ``Catalog of Supply Chain Compromises'' was used as the baseline dataset~\cite{CNCF2023}.
We are not aware of an alternative dataset.
This is a catalog of 69 software supply chain security failures analyzed from news articles and blogs from 1984-2022. 
Each entry describes the failure and its impacts.\footnote{We call these \emph{failures}, rather than ``compromises'', because some cases led to incidents and others were vulnerabilities that were apparently not exploited. See~\cref{sec:Background-FailureStudies}.}
Some examples are in~\cref{table:SampleIncidentsAndResults}.

\subsection{Dimensions of analysis} 
The dimensions of analysis that we replicate and conduct for the software supply chain failures are outlined in ~\cref{table:DIMENSIONS}. 
Additionally, we extend the analysis of the articles in the catalog to explore the capabilities of LLMs at analyzing failures based on data commonly collected to classify and analyze failures \cite{avizienis2004basic}.
We constructed prompts to extract the intent~\cite{avizienis2004basic}, nature~\cite{avizienis2004basic}, impacts~\cite{noauthor_defending_nodate}, and lessons learnt~\cite{melo2021pathology} from the failures.


{
\begin{table*}[ht]
\small
\centering
\caption{
    Dimensions used to analyze the capabilities of LLMs.
    The CNCF database includes ``Type of compromise''.
    Our research team labeled each catalog entry for the next three dimensions.
    The final dimension was assessed via a Likert scale.
  }
\begin{tabularx}{\textwidth}{p{0.2\textwidth}X}
\toprule
\textbf{Dimension} & \textbf{Description} \\
\toprule
Type of compromise & What kind of failure occurred~\cite{CNCF2023}? See~\cref{table:TypesOfAttacks} for types. \\
\midrule
Intent & Was the ``software root cause'' of the failure, accidental or deliberate?~\cite{avizienis2004basic} \\
Nature & Was the failure a vulnerability or an exploit? For exploits, was the actor an insider or outsider?~\cite{avizienis2004basic} \\
Impacts & What kind(s) of impact resulted? The options are taken from~\cite{noauthor_defending_nodate}: (1) Data or financial theft, (2) Disabling networks or systems (3) Monitoring organizations or individuals, (4) Causing physical harm or death (5) All of the above are possible (6) Unknown or unclear.  \\
\midrule
Solutions/learnings & What was the quality of the solutions/learnings from the failure, that the LLM provided ~\cite{melo2021pathology}?  \\
\bottomrule
\end{tabularx}
\label{table:DIMENSIONS}
\end{table*}
}

\subsection{Baseline: Manual Analysis} 

\subsubsection{For RQ1}

The CNCF catalog provides the type of compromise for the failures, stated in \cref{table:SampleIncidentsAndResults}. By manually analyzing the articles, we extend this catalog with three additional dimensions of analysis: intent, nature, and impacts. 

For the dimension of Type of Compromise, the CNCF catalog provides this (analysis conducted by the members of the CNFC organization) and we used their label.
We used existing taxonomies for the dimensions of Intent, Nature, and Impacts, drawing from related works~\cite{avizienis2004basic,noauthor_defending_nodate}.

We had 3 pairs of 2 analysts manually analyze 23 sources per pair for these additional dimensions. They were trained on articles until consistent agreement and definitions were reached.\footnote{The analysts were undergraduate and graduate students in computing, plus one faculty member.} \cref{table:INTERRATER AGREEMENT} shows the inter-rater agreement for these dimensions, measured using Cohen's kappa score. The accuracy for these dimensions was computed in a similar manner. In the case of the ''Impacts'' dimension, we observed a low inter-rater agreement ($\kappa$=\ImpactsAverageKappa).
Given the substantial judgment (or uncertainty) in this dimension, we adopted a ``union'' strategy of accepting the assessment of either rater to determine accuracy. For all other dimensions, disagreements were resolved by the authors.

See~\cref{appendix} for summary distributions of the labels per dimension.

\subsubsection{For RQ2}
For RQ2, we opted not to build a controlled taxonomy of ``lessons learned'' due to the open-ended nature of the prompt.
Instead, we had human raters evaluate the recommendations using a 5-point Likert scale, ranging from "Strongly disagree" to "Strongly agree". The humans rated the LLM's response in relation to the quality of the LLM's response and whether it would mitigate a future attack. 



\begin{table}[ht]
\centering
\caption{
  Inter-rater agreement for the dimensions. The Cohen's kappa ($\kappa$) was calculated for each group (3 groups in total) of raters and then the average $\kappa$ was calculated. 
  }
\begin{tabular}{p{0.4\linewidth}p{0.4\linewidth}}
\hline
\textbf{Dimension} & \textbf{Agreement (Cohen's $\kappa$)} \\
\hline
Type of compromise & Taken as ground truth from the catalog (cf.~\cref{sec:Results-RQ1}) \\
\hline
Intent & \IntentAverageKappa (Group 1- 0.85, Group 2- 1, Group 3- 0.77)\\
\hline
Nature & \NatureAverageKappa (Group 1- 0.60, Group 2- 0.58, Group 3- 0.55)\\
\hline
Impacts & \ImpactsAverageKappa (Group 1- 0.51, Group 2- 0.32, Group 3- 0.20) \\
\hline
\end{tabular}
\label{table:INTERRATER AGREEMENT}
\end{table}



\subsection{Automated approach: LLMs}

\subsubsection{LLM selection}

We used two popular, state-of-the-art LLMs that are publicly available at time of writing (June 2023): OpenAI's ChatGPT model \cite{chatgptcite} and Google's Bard model \cite{bard}.
Their properties are summarized in \cref{table:LLMSummary}.Other large language models are available, \eg~Claude~\cite{anthropic} and Cohere~\cite{cohere}, but GPT and Bard are the most widely used due to their user-friendly interfaces. 


\paragraph{ChatGPT-3.5-turbo, OpenAI's LLM}
GPT-3.5-turbo is a large language model created by OpenAI.
It uses a deep learning method known as transformers.
It is currently one of the most popular and accurate LLMs~\cite{top_6_nlp_language_models_2023}. 
GPT-3.5 uses 175B parameters and is trained on the same datasets used by GPT-3 but with a fine-tuning process called Reinforcement Learning with Human Feedback (RLHF)~\cite{GPTBlog}.

\paragraph{Bard, Google's LLM}
Bard is another popular and accurate LLM created by Google.
Bard also uses transformers.
It uses an optimized version of Language Models for Dialogue Applications (LaMDA) and was pre-trained on a variety publicly available data~\cite{manyika2023bard} including dialogue~\cite{ghahramani2023lamda}. 

{
\begin{table*}[ht]
\small
\centering
\caption{
  Specifications of the LLMs used in the evaluation: GPT-3.5 and Bard.
  GPT's tuning knobs use a 0-1 scale.
  \JD{Adjust the width of columns again?}
}
\begin{tabularx}{\textwidth}{cp{0.12\textwidth}p{0.15\textwidth}cX}
\toprule
\textbf{Model} & \textbf{Cost-to-access} & \textbf{Rate limit} &\textbf{Parameters} &\textbf{Tuning knobs} \\
\toprule
GPT-3.5-turbo-16k \cite{OPENAI2} & Input-\$0.003/1K tokens, Output-\$0.004/1K tokens & 16K tokens per prompt & 175 billion & \textbf{Temperature}: Higher values mean greater model randomness. Default: Unclear. 

\textbf{top\_p}: Nucleus sampling. Model considers the results of tokens with top\_p probability mass. top\_p = 0.2 means to consider only tokens in the top 20\% probability mass. Default: 1. \\
\midrule
Bard~\cite{bard} & Free & Unknown (estimate: 2K tokens per prompt and 50-100 prompts per 9 hours) ~\cite{claude_vs_chatgpt_2023} & 137 billion & \emph{None available to users} \\
\bottomrule
\end{tabularx}
\label{table:LLMSummary}
\end{table*}
}

\subsubsection{Prompt engineering}
A \emph{prompt} is the specific query (instructions or questions) given to an LLM.
The behavior of an LLM varies widely as a result of seemingly minor tweaks to its prompt~\cite{liu2021pretrain}.
Prompt engineering is the process of crafting a prompt for an LLM to increase the quality of its response~\cite{white2023prompt}.

We used prompt engineering to iteratively develop prompts.
We referred to various studies on prompt engineering~\cite{white2023prompt, white2023chatgpt2, openai2023gptbestpractices}.
For each dimension, we refined the prompt by issuing a basic query, then applying each prompt engineering technique in a cumulative sequence until the performance peaked, preserving any changes that improved from the best observed performance.
\cref{table:PROMPT_ENGINEERING_TECHNIQUES} describes our approach using the first dimension, ``Type of Compromise'', as an example.
This prompt engineering phase was conducted on a subset of 20\% of the dataset; we used the most recently published articles.\footnote{We acknowledge that this is a potential source of bias in our results, but did not observe a substantial difference in accuracy between older and newer articles.}
\cref{table:AllPrompts} lists the final version of each prompt.


{
\begin{table*}[ht!]
\small
\centering
\caption{
  Techniques used to improve the prompts, illustrated for the prompt associated with the dimension of type of compromise.
  \emph{`ID'} denotes the order in which the techniques were used. The accuracy column contains the change in accuracy from the previous technique and the final accuracy in brackets. 
  Accuracy was measured over 20\% of the labelled data (we repeatedly analyzed the 14 most recent articles). Prompt 3 was chosen as it had the highest accuracy of 78\%.
}
\begin{tabularx}{\textwidth}{cp{0.175\textwidth}p{0.63\textwidth}c}
\toprule
\textbf{ID} & \textbf{Technique} & \textbf{Prompt} &\textbf{Accuracy (\%)}\\
\toprule

0 & Initial prompt without any techniques &
       "Classify the attack from the following choices
        Choice 1: Dev Tooling
        Choice 2: Negligence
        Choice 3: Publishing Infrastructure
        Choice 4: Source Code
        Choice 5: Trust and Signing
        Choice 6: Malicious Maintainer
        Choice 7: Attack Chaining
 Based on the information provided in the Articles. Article: \{article\} " & \IDZeroTechnique \\
\midrule

1 & Providing context/definitions- adding definitions of the options (improving upon ID: 0) &
        "Classify the attack from the following choices
                Choice 1: Dev Tooling- \textcolor{red}{Occurs when the development machine, SDK, tool chains, or build kit have been exploited. These exploits often result in the introduction of a backdoor by an attacker to own the development environment.}
                
                Choice 2: Negligence- \textcolor{red}{Occurs due to a lack of adherence to best practices. TypoSquatting attacks are a common type of attack associated with negligence, such as when a developer fails to verify the requested dependency name was correct (spelling, name components, glyphs in use, etc).}

                ...
                
                {Based on the information provided in the Article, Article: \{value\}"} & +36 (\IDOneTechnique)\\ 
                
                \midrule
2 &  Reflection Pattern- asking the LLM to explain its answer (improving upon ID: 1) & Adding the sentence \textcolor{red}{"Explain your answer using the given definitions and return the option."} Before passing the article.& +2 (\IDSecondTechnique)\\  
\midrule
3 & Template technique (JSON format) and adding delimiters  (improving upon ID: 2)& Adding \textcolor{red}{"Use JSON format with the keys: 'explanation', 'choice'. Based on the information provided in the Article delimited by triple backticks. Article: \texttt{```\{article\}```"}} in the end. & +7 (\IDThirdTechnique)\\
\midrule
4 & Placement of article- placing the article on top (improving upon ID: 3) &  \textcolor{red}{Based on the information provided in the Article delimited by triple backticks. Article: \texttt{```\{article\}```"}} 
Classify the attack from the following choices & -14 (\IDFourthTechnique) \\
\midrule

5 & The Cognitive Verifier Pattern- asking the LLM to generate addition questions to help it find the correct answer (improving upon ID: 3) & 

" ... 
Choice 7: Attack Chaining- Sometimes a breach may be attributed to multiple lapses, with several compromises chained together to enable the attack. The attack chain may include types of supply chain attacks as defined here. However, catalogued attack chains often include other types of compromise, such as social engineering or a lack of adherence to best practices for securing publicly accessible infrastructure components.

\textcolor{red} {Generate two additional questions that would help you give a more accurate answer. Combine them to produce the final classification. Do not return these questions.}

Explain your answer using the given definitions and return the option. Only return JSON format with the keys: 'explanation', 'option'

Based on the information provided in the Article delimited by triple backticks. Article: ```\{article\}```
                " & -14 (\IDFiveTechnique)\\
\midrule 
6 & Adopt a persona- asking the LLM to look at the article form an expert perspective (improving upon ID: 3) & \textcolor{red}{Act as an software analyst} and classify the attack from the following choices ... & -21 (\IDSixTechnique) \\
\midrule 
7 & Citing evidence- asking for evidence from the text (improving upon ID: 3) &  Explain your answer using the given definitions and return the option. \textcolor{red}{Give evidence from the article to back up your answer.} Use JSON format with the keys: 'explanation', 'option' & -14 (\IDSevenTechnique) \\
\bottomrule
\end{tabularx}
\label{table:PROMPT_ENGINEERING_TECHNIQUES}
\end{table*}
}

\subsection{Experimental Setup}  \label{sec:DataAnalysis}


\subsubsection{Order of prompts}
We prompted LLMs in the order of~\cref{table:AllPrompts}.

\subsubsection{Parameterization of LLMs}

We focused on the two primary adjustable parameters of GPT-3.5, namely ``temperature'' and ``top\_p'', as outlined in \cref{table:LLMSummary}.
According to the literature, when one of the parameters is tuned, the other should be maintained at its default setting \cite{openai2023api}. Our preliminary tests, as shown in \cref{table:PROMPT_ENGINEERING_TECHNIQUES}, were conducted with a temperature of 0 and a default top\_p value of 1.

After finalizing the prompt, we examined the effect of the parameters on accuracy for the ``Type of compromise''.
For this article, accuracy decreased as the temperature increased.
The accuracy was \IDThirdTechnique\% at a temperature of 0, which declined to \Temperatureatpointfive\% at a temperature of 0.5, and further reduced to 50\% at a temperature of 1.
A similar trend was noted for the top\_p parameter.

The optimal performance, with an accuracy of \IDThirdTechnique\%, was achieved with a temperature of 0 and the top\_p parameter at its default value of 1. We retained these parameter settings for the remainder of our analysis. This decision aligns with the guidelines provided in OpenAI's documentation \cite{openai2023api}, which suggests that a lower temperature results in more focused and deterministic responses, a characteristic that is beneficial for article analysis. \footnote{We did not thoroughly test the effect of temperature for RQ2. However, from our testing, GPT either performed similarly or worse with an increase in temperature. Although RQ2 is a more open-ended question, we believe a higher temperature would have led to a response with hallucinations that diverted from the core of the failure.}

\subsubsection{Number of trials}
We noted that the responses of GPT-3.5, configured with Temperature=0, exhibited consistent behavior. 
Consequently, a single trial was conducted to evaluate GPT's accuracy across the dataset. 
Bard's responses were less consistent, but the rate limit was low so we could only conduct one trial.

\subsection{Data Analysis}  \label{sec:DataAnalysis}
We compared the results of the manual analysis against the automated analysis by the LLMs.

For RQ1, we treated each LLM as another analyst and found how accurate it is at classifying various dimensions.
We quantitatively report the LLM's accuracy to measure its correctness for each dimension of analysis.
In cases where the LLM's analysis disagreed with the manual analysis, we examined its justifications.
We qualitatively report some of our observations.

For RQ2, many distinct ``lessons learned'' are possible.
We had analysts review each article and then the recommendations by GPT.
The analysts rated the recommendations on whether the recommendations were appropriate to the article on a 5-point Likert scale:
  ``Strongly disagree'',
  ``Disagree'',
  ``Neither disagree nor agree'',
  ``Agree'',
    and 
 ``Strongly agree''. 
We did not experiment with Bard for this research question due to its rate limits. \\

\section{RESULTS AND ANALYSIS} \label{sec:results}

\subsection{RQ1: How effective are LLMs replicating analysis of SW supply chain failures?} \label{sec:Results-RQ1}

\cref{table: LLM ACCURACY} summarizes the accuracy of GPT and Bard for the type of compromise, intent, nature, and impacts.
GPT consistently outperformed Bard.
We therefore focus our detailed analysis on GPT.

\begin{table}[ht]
\centering
\caption{
  Total accuracy over all the articles for each LLM. 
  }
\begin{tabular}{lcc}
\toprule
\textbf{Dimension} & \textbf{GPT} & \textbf{BARD} \\
\toprule
Type of compromise & \TypeAverageAccuracyGPT & \TypeAverageAccuracyBard\\
Intent & \IntentAverageAccuracyGPT & \IntentAverageAccuracyBard\\
Nature & \NatureAverageAccuracyGPT & \NatureAverageAccuracyBard\\
Impacts & \ImpactsAverageAccuracyGPT & \ImpactsAverageAccuracyBard\\
\bottomrule
\end{tabular}
\label{table: LLM ACCURACY}
\end{table}

\begin{figure}[t]
    \centering 
    \includegraphics[bb=0 0 804 404,width=0.7\textwidth]{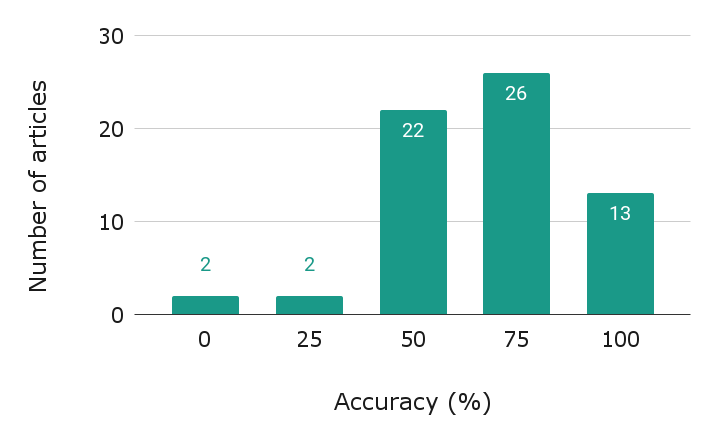}
    \caption{Distribution of the accuracy by articles for GPT. GPT answered 4 questions -- so 5 possible outcomes per case. 
    }
    \label{fig:accuracybyarticles}
\end{figure}

For most articles, GPT performed well on most dimensions.
As depicted in \cref{fig:accuracybyarticles}, GPT demonstrates an accuracy exceeding 75\% (indicating correct responses in three out of four dimensions) in the majority of instances (62\%). 

When the manual raters had higher agreement, GPT tended to agree with them.
GPT had high accuracy in the ``Intent'' and ``Nature'' dimensions, with accuracies of \IntentAverageAccuracyGPT and \NatureAverageAccuracyGPT, respectively.
These dimensions exhibit Cohen's $\kappa$ values of \IntentAverageKappa and \NatureAverageKappa, respectively (\cref{table:INTERRATER AGREEMENT}), demonstrating substantial agreement between the analysts. 
In the ``Impacts'' dimension, the LLM produced an accuracy of \ImpactsAverageAccuracyGPT, as indicated in \cref{table: LLM ACCURACY}. The Cohen's $\kappa$ was also low, at \ImpactsAverageKappa, as shown in \cref{table: SOLUTIONS RATING}.
We conjecture that GPT agrees with analysts when there is a consensus amongst analysts regarding the labeling. 

GPT had trouble when offered multi-answer as an option.
For example, for the ``Impacts'' dimension it could choose from 4 specific impacts, or ``All of the above/Multiple'', or ``Unknown/Unclear''.
In 87\% of the cases, raters chose one of the multi-answer options, while GPT chose one of the specific options.
GPT only selected ``All of the above'' three times and ``Unknown/Unclear'' once.
We conjecture that when GPT was uncertain about the impacts, it opted for the most probable outcome of software supply chain failures in these articles (which focus on IT software).
That option is data and financial theft, which it chose 49 times out of 65.

\begin{figure}[t]
\vspace{0.7cm}

    \centering  
    \includegraphics[bb=0 0 754 404,width=0.7\textwidth]{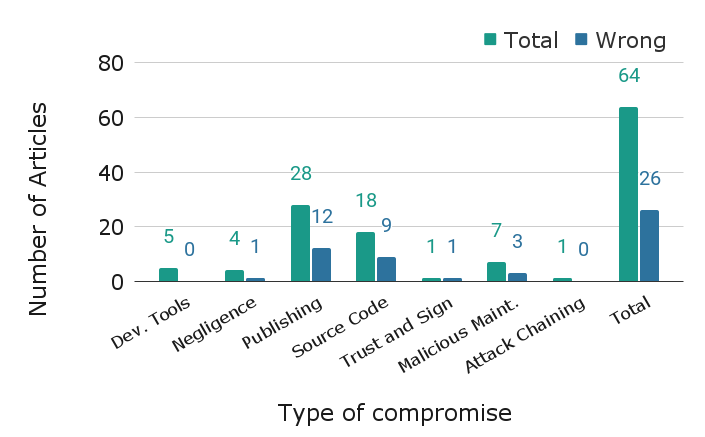}
    \caption{Categorization of articles for the dimension- "Type of Compromise" by GPT. No particular trend is observed. 
    }
    \label{fig:TypeofCompromiseChart}
\end{figure}

We observe that for the articles where the ``Type of compromise'' (ground truth provided by CNCF), we sometimes agreed with GPT over the CNCF.
\cref{fig:TypeofCompromiseChart} represents the distribution of GPT's choice and when they were incorrect according to the CNFC ground truth.
We examined the 14 articles where both the type of compromise and impacts were incorrectly identified.
For these instances, two raters with an inter-rater agreement, $\kappa$ of 0.82 found that most of the time, if they disagreed with CNCF, they concurred with GPT and vice versa.
In the 8 instances where raters disagreed with CNCF, they agreed with GPT 6 times; the same ratio was observed when they disagreed with GPT and agreed with CNCF.
For 2/14 articles they disagreed with both GPT and CNCF.

\vspace{0.7cm}

\vspace{0.7cm}
\vspace{-0.5cm}


\subsection{RQ2: Do LLMs suggest viable mitigation strategies for preventing future failures?} \label{sec:Results-RQ2}
{

To address our second research question, we asked raters to evaluate GPT's proposed solutions/learnings using a 5-point Likert scale. The average ratings are depicted in \cref{table: SOLUTIONS RATING}. The mean score across all three questions is 3.83.
The raters generally held a positive or neutral view of GPT's ``Lessons learned'':
 42\% of the ratings were above 4 (agree), and only 5\% of the ratings fell below 2 (disagree).


For further analysis, we randomly selected two articles where the average score of both the raters $>4$, and two where $<2$.
See~\cref{table:SolutionsandlearningsforGPT} for the full ``Lessons Learned'' for these cases.

\textbf{Factors for strong ratings (average score $\geq$ 4)}.
We believe the LLM demonstrated good performance in these cases due to the depth of the articles. Article 7 \cite{A7} describes the PHP Supply Chain Attack on Pear, and includes technical details of the failures, the exploitation method, and the patch. GPT utilizes the information provided in the blog, combined with its own knowledge, to suggest suitable solutions, \eg
  \emph{"encouraging companies and developers to transition from PEAR to Composer"}.
Article 35 \cite{A35} describes a compromised npm package. It contains technical details of the failure and information on prevention. GPT offers specific solutions,
  such as
  \emph{"encouraging the use of Intrinsic or similar Node.js packages to whitelist and control access to sensitive resources and APIs"}.

\textbf{Factors for weak ratings (average score $\leq$ 2)}.
We believe the LLM demonstrated poor performance in these cases because the articles had few details.
Articles 65 \cite{A65} and 67 \cite{A67} are brief and lack substantial technical details of the failures. Article 67 discusses remote exploitation of a Gentoo server and mentions ongoing forensics. It primarily serves as a notice to users. Article 65 discusses the backdooring of WordPress but provides little information that could inform solutions/learnings. The advice given by GPT is hence generic,
  such as
  \emph{"investigate the incident and address the vulnerability"}
  and
  \emph{"conduct code audits"}.


}


{
\vspace{-0.2cm}
\begin{table}[ht]
\small
\centering
\caption{
  Average rater's rating (Likert scale (1-5/"strongly disagree" to "strongly agree") over all the articles of GPT's response to the solution/learnings prompt.
  }
\begin{tabular}{p{0.8\linewidth}p{0.1\linewidth}}
\toprule
\textbf{Question} & \textbf{Rating} \\
\toprule
Is the advice helpful in general for software supply chain failures? & \FirstQuestionAverage \\
Is the advice related to the specific failure mentioned in the article? & \SecondQuestionAverage\\
Can the advice be used to solve/mitigate the failure mentioned in the article? & \ThirdQuestionAverage\\
\bottomrule
\end{tabular}
\label{table: SOLUTIONS RATING}
\end{table}
\vspace{-0.2cm}
}


\section{DISCUSSION} \label{sec:discussion}

\textbf{Is using LLMs worth it in this context?}
We found that both LLMs in our experiment were capable of simpler forms of analysis, such as distinguishing whether a vulnerability was actually exploited.
However, for more complex questions that require some amount of context or judgment, neither LLM achieved a high level of agreement with the CNCF analysts or our manual raters.
We believe the current generation of off-the-shelf LLMs does not offer a high enough level of agreement with expert judgment to make it a useful assistant 
in this context.
One potential path to improving performance is fine-tuning the LLM using baseline knowledge such as this catalog, and then applying it on future issues~\cite{davis2023reusing}.



\textbf{Will LLMs be a viable alternative to manual analysis in the future?}
In the past few years, OpenAI's GPT models have advanced from simple tasks (GPT-1, GPT-2) to the performance reported here (GPT-3.5).
The recent GPT-4 model is more impressive still~\cite{epson_gpt-1_2023}.
We expect the next generation of LLMs will be suitable aids or replacements for this class of manual analysis. 

\textbf{Future Work.} 
The scope of this analysis could be broadened to encompass additional LLMs, such as Claude~\cite{anthropic} and Cohere~\cite{cohere}.
Additional prompt engineering, and tailoring the prompts per LLM, might improve the accuracy of the results.
Lastly, the analysis could be extended to include a wider range of articles and failures beyond those found in the CNCF catalog~\cite{anandayuvaraj_reflecting_2023,amusuo_reflections_2022}. 

\section{Threats to Validity} \label{sec:Threats}

\noindent\textbf{Internal: }
Prompt engineering was conducted with only one of the LLMs (ChatGPT) utilizing literature from its parent organization (OpenAI); the same prompts were used with the other LLM (BARD). 
The performance of BARD as reported in our study might be misrepresented due to this bias in prompt engineering. 
Additionally, we relied on manual analysis as the ground truth for our evaluation. 
We used multiple raters reaching agreement to mitigate bias. 
We measured an average inter-rater agreement of $\kappa=0.6$, indicating that independent judgments were generally consistent.

Several issues were identified with the catalog and its articles.
(1) Three articles were inaccessible due to broken URLs or PDF formats that were incompatible with LLMs, and were excluded from the analysis~\cite{A26, A59, A55}.
(2) Three articles~\cite{A28, A39, A67} announced a failure, but no analysis --- too little information to answer our RQs.
(3) Some of the CNCF article labels did not match the CNCF taxonomy.
For example, Article 56 \cite{A56} was categorized as a "Fake toolchain", and Article 63 \cite{A63} was labeled as a "Watering-hole attack".
(4) One article~\cite{A61} was not relevant.

Bard's low performance could be due to methodological bias.
Lacking resources on Bard prompt engineering, we used available guidance for GPT.
Bard's limit of 2000 tokens per prompt was below some prompt lengths, potentially reducing accuracy.

\vspace{0.05cm}
\noindent\textbf{External: }
Constructed prompts could be over-fitted to analysis in the catalog. 
Replication of the catalog might not represent failure analysis of incidents in practice. 
Replication of a single catalog might not generalize to all incidents. 

\section{Conclusion}
We evaluate the ability of Large Language Models (LLMs) at characterizing software supply chain failures.
Our study revealed that LLMs are particularly effective when manual analysts are able to reach a consensus on the characteristics of the failure.
In contrast, their performance tends to deteriorate when the agreement among raters is low. 
The quality of the LLMs' outputs also depends on the level of detail provided in the source articles, with more comprehensive articles leading to higher-quality responses.
We conjecture that while LLMs offer a valuable tool for rapidly analyzing large volumes of text, they have not yet reached a stage where they can replace human analysts or manual classification.
Rather than viewing LLMs as a replacement for human input, they should be considered as a supplementary tool that can assist human analysts.
As the depth of detail in postmortems and articles increases, and as LLMs continue to improve, they may evolve into viable analytical resources

\section{Acknowledgments}
OpenAI's ChatGPT model (v4) was used during manuscript preparation.
Prompt: \emph{Can you make the following clearer? ``TEXT SNIPPET''}.
We reviewed answers to ensure it did not change the ideas. 





\clearpage
\balance

\ifBIBLATEX
  \setlength\bibitemsep{0.00\itemsep} 
  \printbibliography
\else
  \bibliographystyle{abbrv}
  \bibliography{refs/sok.bib, refs/jamie-bib.bib, refs/pose-santiago-bibdata.bib, refs/PurdueDualityLab.bib, refs/Chinenye-SoK-bib.bib, refs/paper, refs/references-kelechi.bib}
\fi

\clearpage
\section*{Appendix}\label{appendix}

Table \ref{table:AllPrompts} presents the finalized prompts utilized to query the Language Learning Models (LLMs) across various dimensions. These prompts were derived using a range of prompt engineering techniques, as detailed in Table \ref{table:PROMPT_ENGINEERING_TECHNIQUES}.

{
\begin{table*}[ht]
\small
\centering
\caption{
   The final prompts for each dimension. \DA{Need to fix quotes around article}
  }
\begin{tabularx}{\textwidth}{cX}
\toprule
\textbf{Dimension} & \textbf{Prompt} \\
\toprule
Type of compromise &  Classify the attack from the following choices
                Choice 1: Dev Tooling- This occurs when the development machine, SDK, toolchains, or build kit has been exploited. These exploits often result in the introduction of a backdoor by an attacker to own the development environment.
                Choice 2: Negligence- Occurs due to a lack of adherence to best practices. TypoSquatting attacks are a common type of attack associated with negligence, such as when a developer fails to verify the requested dependency name was correct (spelling, name components, glyphs in use, etc).
                Choice 3: Publishing Infrastructure- Occurs when the integrity or availability of shipment, publishing, or distribution mechanisms and infrastructure are affected. This can result from a number of attacks that permit access to the infrastructure.
                Choice 4: Source Code- Occurs when a source code repository (public or private) is manipulated intentionally by the developer or through a developer or repository credential compromise. Source Code compromise can also occur with intentional introduction of security backdoors and bugs in Open Source code contributions by malicious actors.
                Choice 5: Trust and Signing- Occurs when the signing key used is compromised, resulting in a breach of trust of the software from the open source community or software vendor. This kind of compromise results in the legitimate software being replaced with a malicious, modified version.
                Choice 6: Malicious Maintainer- Occurs when a maintainer, or an entity posing as a maintainer, deliberately injects a vulnerability somewhere in the supply chain or in the source code. This kind of compromise could have great consequences because usually the individual executing the attack is considered trustworthy by many. This category includes attacks from experienced maintainers going rogue, account compromise, and new personas performing an attack soon after they have acquired responsibilities.
                Choice 7: Attack Chaining- Sometimes a breach may be attributed to multiple lapses, with several compromises chained together to enable the attack. The attack chain may include types of supply chain attacks as defined here. However, catalogued attack chains often include other types of compromise, such as social engineering or a lack of adherence to best practices for securing publicly accessible infrastructure components.
                Explain your answer using the given definitions and return the option. Use JSON format with the keys: 'explanation', 'choice'
                Based on the information provided in the Article delimited by triple backticks. Article: ```\{article\}``` \\
\midrule
Intent & Was the root cause of the compromise:		
Option 1: deliberate eg. cyberattack on a system, malicious attackers stealing information		
Option 2: accidental eg. Development incompetence or a bug/vulnerability found		
Explain your reasoning and select an option. Use a JSON format with the keys: 'Explanation', 'option'		
Based on the information provided in the Article delimited by triple backticks. Article: ```\{article\}```	 \\
\midrule
Nature & Was the article about an attack or a vulnerability which was not exploited?		
If it was an attack, who was responsible for the attack?		
Choice 2: Outsider - An attack conducted by an individual or group outside the supply chain, such as a group of terrorists or malicious actors.		
Choice 3: Insider- attack by the developer/someone who was a part of the supply chain		
Choice 4: Unclear		
If it was a vulnerability, return choice 1: Vulnerability		
Explain your reasoning and chose an option.		
Use a JSON format with the key: 'Explanation', 'Option'.		
Based on the information provided in the Article delimited by triple backticks. Article: ```\{article\}```\\

\midrule
Impacts & Classify the attack from the following choices, remember if it is one or more, choose option 5
Option 1: Performing data or financial theft- accessing, extraction, alteration, or destruction of data and/or identity/financial theft.
Option 2: Disabling networks or systems- compromising core functionality, efficiency, or maintainability of the system. Software changes that lead to the product being unusable
Option 3: Monitoring organizations or individuals- keeping track of activities performed by organizations or individuals
Option 4: Causing physical harm or death.
Option 5: All of the above/multiple choices- it is a vulnerability/exploit that can lead to various or all of the impacts from the list.
Option 6: Unknown or unclear
Explain your answer using the given definitions and return the option. Use JSON format with the keys: 'explanation', 'option'
Based on the information provided in the Article delimited by triple backticks. Article: ```\{article\}```\\
\midrule
Solutions/learnings & Suggest the solutions and learning from the failure.		
Use JSON format with the keys: 'Solutions/Learnings'		
Based on the information provided in the Article delimited by triple backticks. Article: ```\{article\}```		\\
\bottomrule
\end{tabularx}
\label{table:AllPrompts}
\end{table*}
}

\cref{table:SolutionsandlearningsforGPT} gives the full set of solutions/learnings proposed by GPT for the four articles discussed in detail in~\cref{sec:Results-RQ2}.

{
\begin{table*}[ht]
\small
\centering
\caption{
   The solutions/learnings outputted by GPT for the four articles mentioned in \cref{sec:Results-RQ2}.
  }
\begin{tabularx}{\textwidth}{cX}
\toprule
\textbf{Article} & \textbf{Solutions/Learnings from GPT} \\
\toprule
Article 7 &  "Implement stronger password reset mechanisms that use secure random number generators instead of weak entropy.",

  "Regularly update and patch software dependencies, such as Archive Tar, to prevent known vulnerabilities.",
  
  "Encourage companies and developers to migrate from PEAR to Composer, which has a more active contributors community and better security practices.",
  
  "Increase security contributions from companies relying on PHP ecosystems like PEAR to identify and fix vulnerabilities.",
  
  "Continuously monitor and scan code repositories for security vulnerabilities using tools like SonarCloud.",
  
  "Improve the validation and sanitization of user input to prevent potential attacks.",
  
  "Educate developers and administrators about secure coding practices and the importance of regularly updating software dependencies.",
  
  "Establish a responsible disclosure process to report vulnerabilities to maintainers and ensure timely patching.",
  
  "Enhance security awareness and training for developers and administrators to identify and mitigate potential threats.",
  
  "Implement access controls and monitoring mechanisms to detect and prevent unauthorized access to critical systems and networks.",
  
  "Regularly review and assess the security of third-party dependencies and libraries used in software development." \\
\midrule
Article 35 &  "Implement stricter security measures when transferring ownership of npm packages to prevent malicious users from gaining control.",

  "Regularly audit and review the dependencies of popular npm packages to identify any potential security vulnerabilities.",
  
  "Encourage open source developers to collaborate and review each other's code to catch any malicious modifications.",
  
  "Consider implementing code signing or verification mechanisms to ensure that the code being uploaded in an npm module is equivalent to the code stored publicly in a git repository.",
  
  "Educate developers about the importance of verifying the integrity and authenticity of dependencies before including them in their projects.",
  
  "Encourage the use of tools that scan npm packages for potential security vulnerabilities, even though they may not be foolproof.",
  
  "Promote the use of Content Security Policy (CSP) to prevent malicious code execution in web applications.",
  
  "Encourage the use of Intrinsic or similar Node.js packages to whitelist and control access to sensitive resources and APIs."\\
\midrule
Article 65 &   "The project should provide more information about the incident, including checksums of good or compromised versions of the software.",

  "The project should investigate and address the vulnerability that allowed the cracker to gain user-level access to the server.",
  
  "The project should take steps to ensure that future WordPress releases are not susceptible to similar security breaches.",
  
  "WordPress users should ask the project serious questions to ensure that they are protected from future incidents.",
  
  "The WordPress developers should conduct code audits to identify and fix any potential backdoors inserted with more care.",
  
  "Free software projects should be aware of the incentive for crackers to insert malware and take appropriate measures to protect their distributions.",
  
  "Projects should focus on securing their processes and servers to prevent attacks and minimize the impact of any potential breaches."\\

\midrule
Article 67 &  "Perform forensic analysis to determine the cause of the remote exploit",

  "Install additional security measures such as intrusion detection systems and file integrity checkers",
  
  "Regularly monitor and analyze server logs for any suspicious activity",
  "Educate users about the importance of running security updates and syncing against trusted servers",
  
  "Consider implementing stronger access controls and authentication mechanisms",
  "Regularly backup critical data to minimize the impact of a compromise",
  "Collaborate with sponsors and infrastructure providers to ensure the security of donated servers",
  
  "Promptly remove compromised servers from rotations and rebuild them after forensic analysis",
  
  "Consider publicly identifying compromised servers to increase transparency and awareness",
  
  "Continuously improve security measures based on lessons learned from incidents"\\
\bottomrule
\end{tabularx}
\label{table:SolutionsandlearningsforGPT}
\end{table*}
}

\begin{figure}[b]
    \centering    \includegraphics[bb=0 0 804 404,width=0.7\textwidth]{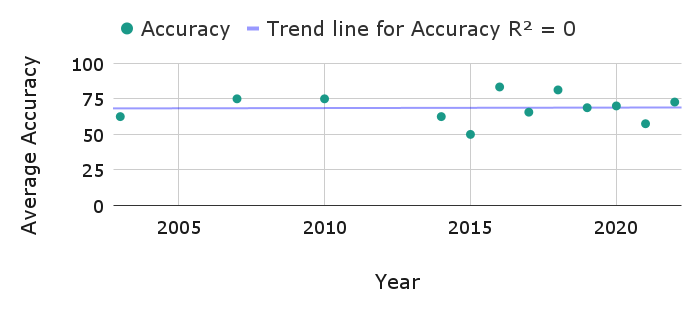}
    \caption{The average accuracy of the articles for all the dimensions over the years. The graph shows no specific trend. 
    }
    \label{fig:AccuracyByYear}
\end{figure}

We wondered whether software supply chain reporting quality has improved over the years.
If this were the case, we would expect to see an increase in LLM performance for newer articles. 
\cref{fig:AccuracyByYear} shows no such trend.

\begin{figure}[b]
\vspace{0.7cm}

    \centering    \includegraphics[bb=0 0 904 404,width=0.7\textwidth]{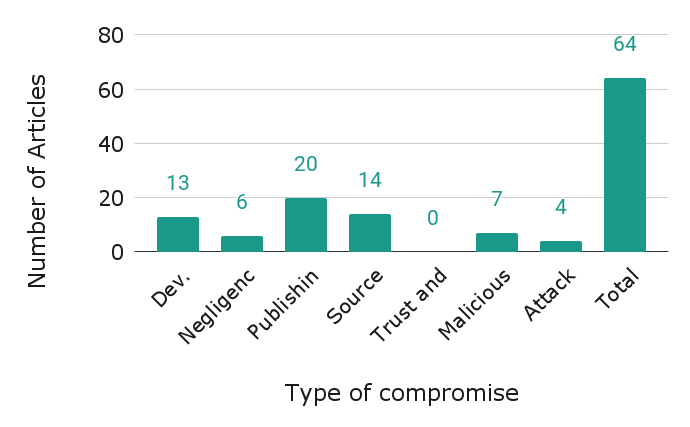}
    \caption{ Categorization of articles for the dimension- "Type
of Compromise" by CNCF catalog.
    }
    \label{fig:TypeGroundTruth}
\end{figure}

\begin{figure}[b]
    \centering    \includegraphics[bb=0 0 904 404,width=0.7\textwidth]{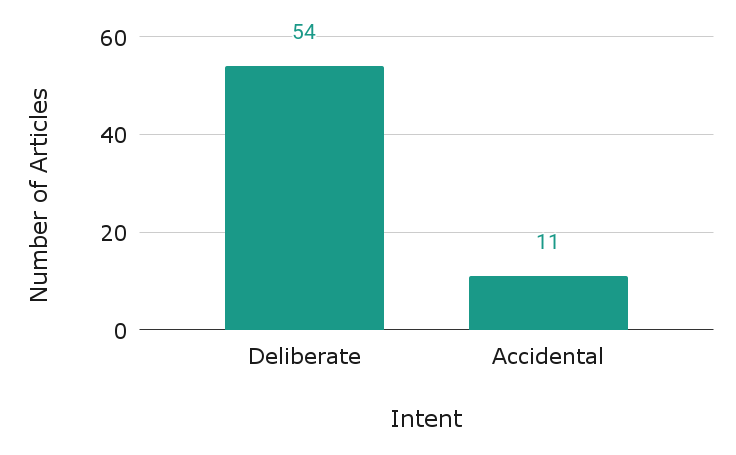}
    \caption{Categorization of articles for the dimension- "Intent" by raters.
    }
    \label{fig:IntentGroundTruth}
\end{figure}

\begin{figure}[b]
\vspace{0.9cm}

    \centering    \includegraphics[bb=0 0 904 404,width=0.7\textwidth]{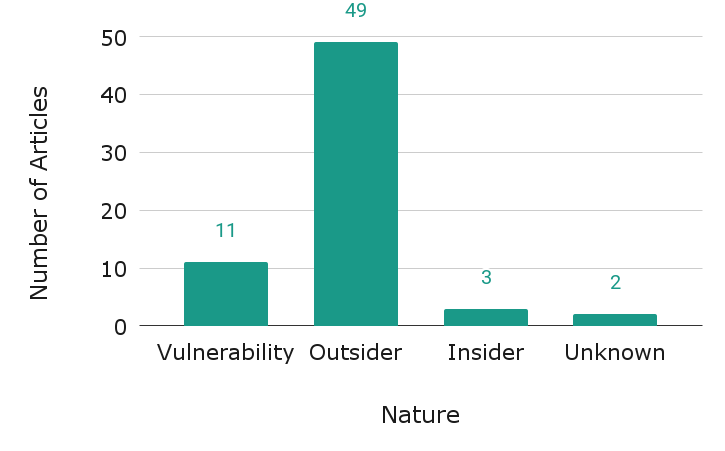}
    \caption{Categorization of articles for the dimension- "Nature" by raters.
    }
    \label{fig:NatureGroundTruth}
\end{figure}

\cref{fig:TypeGroundTruth}, \cref{fig:IntentGroundTruth}, and \cref{fig:NatureGroundTruth} show the ground truth for various dimensions. The ground truth for the dimension ''Impact'' is not presented as the disagreements among the raters were not resolved. In total, there were 65 articles analyzable for the ''Intent'', ''Nature'' and ''Impacts'' dimensions. For ''Type of Compromise'', there were analyzable articles. The failures that were not included were the ones with not functioning URLs and PDF formats, and where the manual labeling of the type of compromise by CNCF was not in the taxonomy. 
\vspace{-1.2cm}
\vspace{0.7cm}

\vspace{0.7cm}

\end{document}

%% file: typesetting.tex
\usepackage{comment}

\usepackage{amsmath,amsfonts}
\usepackage{algorithmic}
\usepackage{graphicx}
\usepackage{textcomp}
\usepackage{xcolor}
\usepackage{booktabs} 
\usepackage{xspace} 
\usepackage[normalem]{ulem}
\usepackage{makecell}
\usepackage{tcolorbox}
\usepackage{enumitem}
\usepackage{siunitx}
\usepackage[vskip=1em,font=itshape,leftmargin=2em,rightmargin=2em]{quoting}

\usepackage{soul}

\ifDEBUG
    \newcommand{\JD}[1]{\textcolor{purple}{[JD:#1]}}
    \newcommand{\TRS}[1]{\textcolor{olive}{[Taylor:#1]}}
    \newcommand{\DA}[1]{\textcolor{blue}{[DA:#1]}}
    \newcommand{\TS}[1]{\textcolor{red}{[TS:#1]}}
    \newcommand{\TODO}[1]{\hl{#1}}
\else
    \newcommand{\JD}[1]{}
    \newcommand{\TRS}[1]{}
    \newcommand{\DA}[1]{}
    \newcommand{\TS}[1]{}
    
    \newcommand{\TODO}[1]{}
\fi

\newif\ifSPACEHACK
\SPACEHACKfalse
    \usepackage{etoolbox}
    \makeatletter
    \patchcmd{\ttlh@hang}{\parindent\z@}{\parindent\z@\leavevmode}{}{}
    \patchcmd{\ttlh@hang}{\noindent}{}{}{}
    \makeatother

\ifSPACEHACK
    \titlespacing*\section{0pt}{1pt plus 1pt minus 1pt}{1pt plus 1.5pt minus 1.5pt}
    \titlespacing*\subsection{0pt}{1pt plus 1.5pt minus 1.5pt}{1pt plus 1.5pt minus 1.5pt}
    \titlespacing*\subsubsection{0pt}{2pt plus 1pt minus 1pt}{1pt plus 1.5pt minus 1.5pt}
    \titlespacing*\paragraph{0pt}{1pt plus 1.5pt minus 1.5pt}{1pt plus 1.5pt minus 1.5pt}
\fi
\usepackage{balance} 
\usepackage{subcaption}

\usepackage{amsmath}
\usepackage{cleveref}

\crefformat{section}{\S#2#1#3}
\crefname{figure}{Figure}{Figures}
\crefname{appendix}{Appendix}{Appendices}
\crefname{table}{Table}{Tables}
\crefname{algorithm}{Algorithm}{Algorithms}
\crefname{listing}{Listing}{Listings}
\crefname{theorem}{Theorem}{Theorems}
\crefname{thm}{Theorem}{Theorems}
\crefname{lemma}{Lemma}{Lemmata}
\crefname{equation}{Eqt.}{Eqts.}
\crefformat{Grammar}{Grammar #1}

\usepackage{enumitem}
\usepackage{booktabs}

\newcommand{\eg}{\textit{e.g.,} }
\newcommand{\etal}{\textit{et al.}\xspace}



%% file: data/data.tex
\newcommand{\IntentAverageKappa}{0.87\xspace}
\newcommand{\NatureAverageKappa}{0.58\xspace}
\newcommand{\ImpactsAverageKappa}{0.34\xspace}

\newcommand{\TypeAverageAccuracyGPT}{59\%\xspace}
\newcommand{\IntentAverageAccuracyGPT}{88\%\xspace}
\newcommand{\NatureAverageAccuracyGPT}{74\%\xspace}
\newcommand{\ImpactsAverageAccuracyGPT}{52\%\xspace}

\newcommand{\TypeAverageAccuracyBard}{28\%\xspace}
\newcommand{\IntentAverageAccuracyBard}{88\%\xspace}
\newcommand{\NatureAverageAccuracyBard}{69\%\xspace}
\newcommand{\ImpactsAverageAccuracyBard}{45\%\xspace}

\newcommand{\FirstQuestionAverage}{3.72\xspace}
\newcommand{\SecondQuestionAverage}{4.15\xspace}
\newcommand{\ThirdQuestionAverage}{3.62\xspace}

\newcommand{\IDZeroTechnique}{33}
\newcommand{\IDOneTechnique}{69}
\newcommand{\IDSecondTechnique}{71}
\newcommand{\IDThirdTechnique}{78}
\newcommand{\IDFourthTechnique}{64}
\newcommand{\IDFiveTechnique}{64}
\newcommand{\IDSixTechnique}{57}
\newcommand{\IDSevenTechnique}{64}

\newcommand{\Temperatureatpointfive}{64}